\documentstyle[12pt,aasms4]{article}

\def\gtsim{\lower.5ex\hbox{$\; \buildrel > \over \sim \;$}}
\def\ltsim{\lower.5ex\hbox{$\; \buildrel < \over \sim \;$}}

\begin{document} 

\tighten \title{THE {\sc WARPS} X-RAY SURVEY OF GALAXIES, GROUPS AND
CLUSTERS -- I. Method and first results} \author{C.A. Scharf
\altaffilmark{1}, L.R. Jones \altaffilmark{1}}
\affil{Lab. for High Energy Astrophysics, Code 660,
NASA/Goddard, Greenbelt MD 20771, USA, email: caleb@amadeus.gsfc.nasa.gov, lrj@clopper.gsfc.nasa.gov}
\author{H. Ebeling}
\affil{Institute of
Astronomy, Madingley Road, Cambridge CB3 OHA, UK {\em and} present address\/ Institute
for Astronomy, 2680 Woodlawn Dr, Honolulu HI 96822, USA, email: ebeling@marvin.ifa.hawaii.edu}
\author{E. Perlman \altaffilmark{2}}
\affil{Lab. for High Energy Astrophysics, Code 660,
NASA/Goddard, Greenbelt MD 20771, USA, email: perlman@baley.gsfc.nasa.gov}
\author{M. Malkan}
\affil{Dept. of Astronomy, UCLA, Los Angeles, CA 90024, USA, email: malkan@bonnie.astro.ucla.edu}
\and
\author{G. Wegner}
\affil{Dept. of 
Physics \& Astronomy, Dartmouth College, 6127 Wilder Lab., Hanover, NH 
03755, USA, email: wegner@kayz.dartmouth.edu}

 \altaffiltext{1}{NRC Research Associate}
  \altaffiltext{2}{USRA}

\begin{abstract} We have embarked on a survey of
 {\sl ROSAT} PSPC archival data with the aim of detecting {\em all}
significant surface brightness enhancements due to sources in the
innermost $R \leq 15$ arcmin of the PSPC field of view  in the energy
band $0.5-2.0$ keV. This project is part of the Wide Angle ROSAT
Pointed Survey (WARPS) and is designed primarily to measure the low
luminosity, high redshift, X-ray luminosity function of galaxy clusters
and groups.
 The approach we have chosen for source detection [Voronoi Tessellation
 and Percolation (VTP)] represents a significant advance over
conventional methods and is particularly suited for the detection and
accurate quantification of extended and/or low surface brightness
emission which could otherwise be missed or wrongly interpreted. We
also use energy dependent exposure maps
 to estimate the  fluxes of  sources which can amount to corrections of
 as much as 15\%.  In an extensive optical follow-up programme we are
 identifying galaxies, groups and clusters at redshifts ranging
from $z \sim 0.1$ to $z \sim 0.7$.

In this paper we present our method and its calibration using simulated
and real data. We present first results for an initial 91 fields ($
17.2 $ deg$^{2}$) at detected fluxes $> 3.5 \times 10^{-14}$ erg
s$^{-1}$ cm$^{-2}$
 (the WARPS-I survey). We find the sky density of extended objects to
 be in the range 2.8 to 4.0 $(\pm 0.4)$ deg$^{-2}$.  A comparison with
a point source detection algorithm demonstrates that our VTP approach
typically finds 1-2 more objects deg$^{-2}$ to this detected flux limit,
suggesting that the conventional method fails to detect a significant
fraction of extended objects. The surface brightness limit of the WARPS
cluster survey is $\sim 1\times 10^{-15}$
 erg sec$^{-1}$ cm$^{-2}$ arcmin$^{-2}$, approximately 6 times lower
than the Extended Medium Sensitivity Survey (EMSS).  The WARPS
LogN-LogS (which currently represents a lower limit) shows a
significant excess over previous measurements for S$\gtsim 8\times
10^{-14}$ erg sec$^{-1}$ cm$^{-2}$. We attribute this mainly to a
larger measured flux from extended sources as well as new detections of
low surface brightness systems in the WARPS.

\end{abstract}


\section{Introduction}

Hierarchical models of structure formation in the Universe (such as
Cold Dark Matter (CDM) cosmologies) predict the basic evolutionary
properties of gravitationally bound aggregates. Clusters of galaxies
are the largest such objects to have already decoupled from the Hubble
expansion and therefore offer a unique opportunity to evaluate the
fundamental properties of the Universe.  The dominant baryonic mass
component in galaxy clusters is observed to be in the form of hot
($10^{7} - 10^{8}$ K) intra-cluster gas ($\sim 90$\% of the luminous
mass) emitting radiation through thermal bremsstrahlung with some
contribution from thermal line emission. We therefore expect the X-ray
luminosity to be positively correlated with the system mass (assuming a
constant baryon mass fraction, see e.g. \markcite{White93}White et al
1993). Since the measurement of cluster X-ray luminosity is relatively
easy, it follows that to investigate the nature of clusters and their
formation we should choose a primary selection method based on X-ray
observations.

In the past, selection effects have beset optical surveys because of
the projection of background and foreground galaxies on the cluster,
sometimes leading to false identifications (e.g.\markcite{Frenk90}
Frenk et al 1990), particularly for poor clusters at high redshift,
where the contrast with the background galaxy surface density is low.
Even with a knowledge of the line-of-sight dynamics of a system it is
only possible to assign probabilities for individual objects to belong
(gravitationally) to a cluster or group of galaxies.  X-ray surveys
suffer less from such problems. Since the X-ray surface brightness is
proportional to the square of the density of the hot gas within the
cluster, the contrast with respect to the unresolved X-ray background
is high. X-ray observations also provide information about the
potential size and shape leading to much less ambiguous identification
of real physical systems \footnote{ With sufficiently good spectral
resolution and signal-to-noise (such as that provided by ASCA,
\markcite{Tanak94} Tanaka et al 1994) one could also determine the
velocity distance {\em of the cluster potential} directly, using (for
example) the Fe K-$\alpha$  line, rather than taking the mean over
individual galaxies which contains shot noise.}.  Well selected
catalogues of galaxy clusters can then be used to investigate the
evolution of the cluster X-ray luminosity function, the morphology and
substructure of clusters, the temperature evolution of the gas
component, and the   evolution of the optically observed galaxy
population.

What do we expect the X-ray cluster evolution to be in a hierarchical
model ? It would be fair to say that both the observational and
theoretical situation are still uncertain. Modeling of the
gravitational instability and gas hydrodynamics of clusters has been
pursued with semi-analytic methods \markcite{Kaise91} (Kaiser 1991) and
numerical simulation (\markcite{Evrar90} Evrard
1990,\markcite{Bryan94a} Bryan et al 1994a,\markcite{Cen94} Cen \&
Ostriker 1994,\markcite{Kang94} Kang et al 1994 ). From such studies it
is evident that the evolutionary properties of clusters are sensitive to
both the assumed 'internal' cluster physics (e.g. the presence or
absence of cooling flows or feedback mechanisms) as well as the
'external' influence of the underlying cosmological model (e.g.
Freidmann model parameters and the power
 spectrum of mass fluctuations).  However, in the case of relatively
simple 'internal' cluster physics (e.g. shock heating during
gravitational collapse as a dominant mechanism) and a standard CDM
cosmology there is a prediction of  negative evolution (i.e. a
decrease with redshift) in the number density of the {\em most
luminous} clusters (with $L_{x}> 5\times 10^{44}$ erg s$^{-1}$ ).

Current surveys broadly confirm  this scenario. The EMSS cluster sample
of \markcite{Henry92} Henry et al (1992) showed evidence (at the
3$\sigma$ level) of negative evolution in the cluster X-ray luminosity
function (XLF), with fewer high luminosity clusters at redshifts
0.3$<$z$<$0.6 than at 0.14$<$z$<$0.2.  At lower redshifts, the XBACS
and BCS cluster samples recently compiled by \markcite{Ebeli96a}
\markcite{Ebeli96b}Ebeling et al (1996a,b), based on ROSAT All-Sky
Survey data, show little evolution of the Schechter function XLF within
a redshift of $z=0.3$ (although the small amplitude of negative
evolution actually measured is statistically highly significant).  The
earlier detection of evolution at $z<0.2$ by \markcite{Edge90} Edge et
al (1990) has been shown to be due to an unfortunate sampling of
clusters at $z=0.1$ to 0.15, rather than to evolution
(\markcite{Ebeli95} Ebeling et al 1995). At high redshifts
(0.2$<z<$0.6) and lower luminosities, \markcite{Casta95} Castander et
al (1995) find evidence for strong negative evolution. However, this
result is based on only 13 clusters, and the sample completeness is
unclear (see also  Section 5 below).  Existing X-ray selected surveys
have not probed the low-luminosity regime beyond  z$\sim$0.15, except
for the small sample of \markcite{Casta95} Castander et al (1995).

The slope of the cluster XLF steepened with redshift in the EMSS survey
of \markcite{Henry92} Henry et al (1992) such that at lower
luminosities (around $10^{44}$ erg s$^{-1}$) no evolution was
required.  If the steep slope of the EMSS high redshift XLF is
extrapolated to still lower luminosities (L$_X<10^{44}$ erg s$^{-1}$),
positive evolution of low luminosity clusters is predicted, with large
numbers ($\sim$8 deg$^{-2}$) of z$>$0.2 clusters at low X-ray fluxes
($>3 \times 10^{-14}$ erg cm$^{-2}$ s$^{-1}$).  Accurate measurements
of the high redshift XLF would allow the {\it form} of the XLF
evolution to be determined via the position of the Schechter function
break.  This would help discriminate between luminosity and density
evolution, and discriminate between different hierarchical models, e.g
those including a different mix of fundamental particles (e.g.
\markcite{Bryan94b} Bryan et al 1994b), a flat power spectrum of the
initial fluctuations (\markcite{Henry92} Henry et al 1992) and
reheating of the intracluster gas at high redshifts (\markcite{Kaise91}
Kaiser 1991).  The WARPS cluster survey was designed to make this
measurement (\markcite{Jones96}Jones et al 1996).

To ensure the completeness of the WARPS sample a source detection
method (Voronoi Tessellation and Percolation, VTP) is used which does
not discriminate  {\em against} extended sources of low surface
brightness and/or irregular morphology. This is  especially important
because of the evolution that might be expected in properties such as
core radius, temperature and morphology. Indeed, cluster morphology has
been shown to be a potentially important cosmological discriminator
(\markcite{Evrar93} Evrard et al 1993, \markcite{Crone94} Crone, Evrard \& Richstone 1994).
While luminous clusters at high redshift ($z>0.4$) are likely to be
 detected by standard point source search algorithms, those at lower
 $z$ and lower luminosity may be missed  altogether by such methods
 (Section 5) or removed from a flux limited survey due to an
 incorrectly low flux detection.  An analogous situation exists in
 optically selected galaxy catalogues - a magnitude selected sample with 
 a high surface brightness detection threshold
 will be incomplete in diameter and vice versa. Such selection effects
 will bias conventional catalogues against (for example) extended low
 surface brightness objects and cause fluxes to be incorrectly
 measured.

In this paper we describe the methods used in the WARPS survey and
present its first results.  This work is arranged as follows: in
Section 2 we present a brief overview of the WARPS survey and the ROSAT
PSPC  data archive,
 in Section 3, 3.1, 3.2 and 3.3 we describe the VTP method,  the
construction of a flux limited catalogue and discuss the accuracy of
flux corrections and extended source detection. Section 4 presents the
calibration of the survey sky coverage. Section 5 presents a comparison
with the results of a conventional detection algorithm. In  Section 6
we discuss the current results and in  Section 6.1 present a LogN-LogS
for WARPS-I . In Section 7 we summarize this paper and present
conclusions.

\section{The WARPS survey and the {\sl ROSAT} data}

The WARPS survey is based on archival ROSAT  Position Sensitive
Proportional Counter (PSPC) X-ray data. The aim of the WARPS cluster
survey is to obtain a well-calibrated, complete sample of all sources
which emit X-rays from hot gas trapped in a gravitational potential,
from  single galaxies to rich clusters.  The X-ray images are searched
for serendipitous sources using a surface brightness limit, and those
sources with detected flux $> 3.5 \times 10^{-14}$ erg s$^{-1}$
cm$^{-2}$ (0.5-2 keV) are classified as extended or point-like.  In an
ongoing optical follow-up programme of the extended sources and
selected point-like sources (specifically those with galaxy
counterparts) we are obtaining both imaging (from archived plate data
and deeper CCD data) and spectroscopy data. The followup procedure has
been designed to minimize incompleteness and misidentifications of the
X-ray source candidates (catalogue in preparation). The spectroscopic
redshifts of cluster galaxies will then allow us to determine the XLF
and measure its evolution.

The ROSAT PSPC provides a 2 degree diameter field of view with an
energy range of $0.1 - 2.4$ keV and a modest energy resolution. The
relative  positional accuracy of the photon coordinates in the
instrument plane is $0.5$ arcsec. The shape and size of the
instrumental point spread point spread function (PSF) depends on both
photon energy and off-axis angle.  For a photon energy of 1 keV the PSF
has a full width half maximum (FWHM) of 25 arcsec on axis and increases
in size with off-axis angle to a FWHM of 58 arcsec  at 15 arcmin off
axis (\markcite{Hasin93b} Hasinger et al 1993b). We limit our survey to
sources within this off-axis angle.
 We use the 0.5-2 keV band to detect sources rather than the full
 0.1-2.4 keV PSPC band in order to (a) reduce the contribution to the
 background from gas in our Galaxy at $\sim 10^{6}$ K, (b)  minimize
 the size of the PSF, and (c) to maximize sensitivity to hard sources
 such as clusters of galaxies.

There are currently 4768 PSPC fields in the ROSAT archive. Some 1400
fields have exposure times in excess of 8 ksec and we choose this as
one criterion for selecting fields our survey. Our field selection is
then based on the following general criteria. We first consider fields
in which the primary targets  are stars or active galactic nuclei (AGN)
and which contain no very bright optical objects (typically bright
stars which then make optical follow-up difficult).  We avoid pointings
whose targets are bright clusters since the latter dominate the inner
15 arcmin of the PSPC's field of view and make serendipitous detections
of other objects impossible. This also ensures that our cluster
selection is not affected by the known angular correlation between
clusters. In this initial survey we have also chosen high galactic
latitude ($|b|>20^{\circ}$) fields. The observed distribution of
Galactic equivalent column densities of neutral hydrogen (all $<
10^{21}$ cm$^{-2}$) implies that there should be virtually no
absorption in the 0.5-2 keV band used here. The corrections to the
measured X-ray fluxes are $<2$\% on all fields and  are therefore well
within the total flux uncertainties.

 In those fields selected we exclude
objects within the inner 3 arcmin radius of the field centre since
these normally constitute the original targets.

In Figure 1 the positions of the 91 ROSAT fields selected in this
initial survey (WARPS-I) are plotted in Galactic coordinates. The size
of the points is proportional to the exposure time which ranges from
$\sim 8000$ seconds to $\sim 48,000$ seconds, with most fields having
exposures $\ge 10,000$ seconds.

\section{The VTP method}

The VTP method of Ebeling (\markcite{Ebeli93}Ebeling 1993,
\markcite{Ebeli93} Ebeling \& Wiedenmann 1993) is a general method for
the detection of non-Poissonian structure in a distribution of points.
In the case of a distribution of photons it will detect all regions of
enhanced surface brightness (surface density) relative to the
Poissonian expectation.
 While the ROSAT PSPC photons are in fact registered on a finite grid
 made from 0.5 arcsec `pixels', we can treat the observed photon
 distribution as unbinned since, at the exposure times typically
 attained in the pointings, this grid is well sampled only at the
 positions of the very brightest sources.  The VTP method can be
summarized as follows. First, for a raw photon distribution (in this
case the inner 15 arcmin radius of the {\sl ROSAT} PSPC and a `buffer
zone' of photons to 18 arcmin) the unique Voronoi tessellation is
determined (when rare multiple photon counts occur at a grid position,
these are flagged and this information is used in the percolation
described below).  Each photon defines the centre of a cell polygon
whose sides form the perpendicular bisectors of the non-crossing
vectors joining the nearest neighbour photons which, in turn,
represent the vertices of the equivalent Delauney triangulation. These
photon cells form the Voronoi tessellation of the field. Since most
cells contain only one photon, the surface brightness associated with
this photon equals the inverse of the product of cell area and local
exposure time.  In Figure 2 the Voronoi tessellation or tiling is
shown for a typical ROSAT PSPC field. The sources in this field are
apparent to the eye as the clusters of  small cells.

The PSPC field experiences non-uniform exposure which can vary by as
much as 15-20\% from the field center to 15 arcmin radius off axis, and
by as much as 10-15\% between different fields (according to the
spacecraft parameters at a given time). Extended sources will therefore
also experience non-uniform exposure across their projected surface
(for example $\sim 5$\% across a 3 arcmin source). To correct for this,
we generate exposure maps for each pointing  (with $15'' \times 15''$
resolution pixels) using an algorithm based on the detailed work of
\markcite{Snowd92} Snowden et al (1992). For greater accuracy, exposure
maps are constructed in two energy bands ( 0.5-0.9 keV and 0.9-2 keV)
and the appropriate map is used to yield an exposure value {\em for
each photon}. Although the mean difference between the global broad
band (0.1-2 keV) exposure map and those used here
 is only $\sim 5$\%, the variation between individual maps can be as
 much as 15\%.  The final improvement on estimated source fluxes
(especially for extended sources) obtained by using the correct
exposure maps can therefore be as much as 10-15\% when compared to
fluxes with only a uniform exposure correction.

The cumulative distribution of the inverse areas of the Voronoi cells
can then be compared with that expected from a random (Poisson)
distribution and a cutoff can be determined which defines the global
background count for that field.  A spatial percolation algorithm is
then run on the cells with areas smaller than a given threshold above
the background, grouping them according to the excess above the
background density and forming sources. The latter two steps are
performed in an iterative fashion so that as sources are detected the
background estimate is revised and the source groupings redetermined.
Typically this requires $\sim 6$ iterations. Finally, the minimum
number of photons required for a true source is calculated (such that
we expect 1 fake source in the total survey area) to eliminate
background fluctuations. The background level for a given field is then
 simply the mean surface brightness ($\sigma_{back}$) calculated from all non-source photons.

 For each source a series of  parameters are then obtained by VTP. In
 our analysis we use the following: the source position (determined as a
 weighted sum over photons), the detected source count rate (corrected
 for the background count rate), the detected source area, the minimal
 and maximal moments of inertia of the photon distribution, an estimate
 of the background count, and the probability of the source being a
 statistical fluctuation (calculated as the probability of the Poisson background producing a fluctuation of that number of  
 photons with local surface brightness above the detection threshold). In addition, the full set of photons
 associated with each source is stored.

\subsection{Source deblending}

We perform VTP three times for each field using different surface
brightness thresholds (denoted as factors of the background surface 
brightness: 1.0, 1.3 and 1.7; see Section 4). 
This allows us to distinguish real single sources from those composed of
blends of several sources (point-like or extended) and reduces any
uncertainties in source identification due to positive background
fluctuations which become grouped with source photons. 

 The first task
in the survey (once VTP has been run) is therefore to perform the
deblending and threshold selections. For this purpose we combine an
automated deblending procedure with visual inspection of all fields.
The deblending algorithm is run over all VTP results. It includes all
sources which match the following criteria:  \begin{itemize} \item
Sources at threshold 1.0 which do not deblend at higher thresholds, lie
 within 15 arcmin of the field centre and typically outside the innermost
3 arcmin radius (to avoid target sources) and with an observed count
rate of $>3\times 10^{-3}$ ct/s (corresponding to an observed flux
limit of $\simeq 3.5 \times 10^{-14}$ erg s$^{-1}$ cm$^{-2}$) \item
Sources matching the above position and count rate criteria  at
threshold 1.3 which were parts of blends at 1.0 but do not deblend at
threshold 1.7 \item Sources matching the position and count rate
criteria at threshold 1.7 which were parts of blends at 1.3

\end{itemize}

This first run of the deblending algorithm  provides a list of
candidates which is used in a visual inspection of each field. The
choices which may then be made are to alter the detection threshold
used for either the entire field or for individual objects. The
rationale for this is that sometimes objects detected at threshold 1.0
will include (because of the high sensitivity of VTP) photons which are
clearly positive noise and act to bias the area estimate of the source
(i.e. spurious `tails' of very low surface brightness which become
associated with a source). Increasing the threshold removes this noise
and typically only removes $<10$\% of the source photons (which will be
recovered in the flux correction step detailed below). The threshold
used for each source is recorded and used in the correction from
detected to total flux as described in Section 3.2.

 Figures 3 and 4 demonstrate the differences between the lowest and
 highest threshold results for the field in Figure 2.  Those photons
 identified as belonging to sources by VTP are plotted in heavy type.
 Sources typically account for 10-20\% of the photons in the lowest
 threshold, less in the highest threshold.
 The sources  numbered 2 and 6 in Figure 3 are strong candidates for
 blends of more than one source when observed at the lowest threshold.
 In Figure 4 it is apparent that these sources have been deblended at
 the higher (1.7) threshold.

Clearly there may be cases of real physical systems which contain
structure  that becomes deblended. However, a visual inspection of all
91 fields revealed only 2 cases where deblending had split up what was
probably a single extended object (which had a flux above the survey
flux limit) into separate components which fell below the flux limit.
These 2 objects were placed back into the survey sample. In addition,
since we obtain optical identification for {\em all} sources
(both extended and point-like) likely to be clusters, groups or normal
galaxies, we expect to be able to catch such cases, should they occur.
As an additional aid in deblending, the WGACAT point source detections
(\markcite{White94} White, Giommi \& Angelini 1994) are plotted as
vertical arrows in Figures 3 and 4.  Note that, since the WGACAT
sources were detected in the broader 0.24-2 keV band, they cannot be
compared directly to the VTP detections; a detailed comparison between
VTP and conventional detection algorithms is made in  Section 5.

\subsection{Flux determinations}

The final list of deblended sources and their respective VTP
parameters are then passed to an algorithm which estimates the true
flux of the sources. In order to do this, the general nature of VTP must
be abandoned and assumptions made about the nature of the sources. We
assume that sources are either intrinsically point-like or extended with
a surface brightness distribution following a King profile:

\begin{equation} \sigma (r)= \sigma_{0} \left[ 1+(r/r_{c})^{2} \right]
^{-3\beta +1/2} \;\;\;, \end{equation} where $\sigma(r)$ is the
projected surface brightness at distance $r$ from the
 source centre, $\sigma_{0}$ is the central surface brightness, $r_{c}$
 is the core radius, and $\beta$ has a value of 2/3 (\markcite{Jones84}
 Jones \& Forman 1984).  In modeling the PSF we follow
 \markcite{Hasin93b} Hasinger et al. (1993b).

The VTP algorithm returns estimates of the mean surface brightness of
the background, a background corrected estimate of the observed count
rate for each detected source, and the area of the source above the
surface brightness threshold. From the area we obtain an area
equivalent source radius ($r_{VTP}=\sqrt{A_{VTP}/\pi}$). Two flux
correction factors are then computed for each source; the first assumes
it is point-like. The second assumes it is extended with a King profile
(Equation 1) and
 with a knowledge of the surface brightness threshold we can
numerically obtain an equivalent core radius ($r_{c}$) (corrected for
PSF) and  the central surface brightness ($\sigma_{0}$).  The true
(integral) source count rate (i.e. including the undetected flux below
the threshold) is then simply obtained as:

\begin{eqnarray} s_{true} = 2\pi \int_{0}^{\infty} \sigma (r) r dr =
\frac{\pi \sigma_{0} r_{c}^{2}}{3(\beta - 1/2)} \;\;\; .
\end{eqnarray}

For simulated data (using idealized surface brightness profiles, such
as the King profile) the VTP corrected fluxes are accurate for high
signal-to-noise sources (described in Section 3.3 and Figure 6). For
low signal-to-noise sources the uncertainty in the corrected flux is
larger (e.g. Figure 5). Full details of this procedure can be found in
\markcite{Ebeli96b} Ebeling et al (1996b).

 An object is then classified as extended if the ratio of the
correction factors $f=s_{true}/s_{detected}$ for a King profile and a
point source, respectively, lies above a critical value. The true count
rate is then obtained by multiplying the detected count rate with the
appropriate correction factor.
  Since it is only integral properties that are used, the results are
much more stable to deviations of objects from the assumed ideal than
in the fitting procedures used by conventional detection algorithms.

\subsection{Flux corrections and extents}

To test and calibrate the flux correction method described above we
have used Monte Carlo simulations of PSPC fields. Sources are simulated
with point-like or King profile surface brightness, convolved with the
instrumental  PSF (for a nominal photon energy of 1 keV) and added to
a  representative background of Poissonian
 noise. An ensemble of sources are made for 40 sets of typical intrinsic
 parameters (extent, flux, off-axis angle) to determine the expected
scatter in the VTP estimates. To further mimic the selection of the
real survey data we inspect the simulated sources to choose the
appropriate surface brightness threshold to use which eliminates false
positive noise wings.

In Figures 5(a,b) and 6(a,b)  results are presented across a large
range of source parameters, from extents of $\sim 0$ arcsec to 1.5
arcmin and effective fluxes (0.5-2 keV) of $\sim 6\times 10^{-14}$ to
$\sim 3.5 \times 10^{-13}$erg s$^{-1}$cm $^{-2}$ and for on-axis (circular) and off-axis (triangular) sources. In all plots the
ratio of the raw, detected count rates to the true count rate and the
ratio of the corrected (see above) count rates to the true count rate
are plotted on the y-axes. Multiple points at fixed intrinsic extents 
 (or fluxes) have a range of intrinsic fluxes (or extents). 

The effective VTP detected signal-to-noise (as defined in Section 4
below) of a source is a good indicator of the reliability of any flux
estimate. The median signal-to-noise of the real X-ray source
detections is $s/n=8$, and we use this to divide the simulation results
for presentation.  The effect of noise is apparent in the larger
scatter seen in flux estimates for the low signal-to-noise ($s/n< 8$)
sources plotted in Figure 5(a,b) when compared to those of the high
signal-to-noise objects in Figure 6(a,b).  For sources of medium to
small extent (i.e. $<60$ arcsec) with signal-to-noise $> 8$ we can
recover the true flux to within 5-10\% over the  flux range shown
here.  For sources of larger extent (of which we might expect to see
 very few in our survey) the fluxes are systematically underestimated,
 as expected since more of the flux now lies below our surface
 brightness limits.  For the low signal-to-noise sources ($s/n < 8$)
the same general trends are present but dominated by the scatter due to
noise. For small extents ($< 30$ arcsec) we can still recover the true
flux of the faintest objects to within 10-20\%. To summarize; all
recovered count rates are within 1-$\sigma$ of the nominal count rates,
except for highly extended ($> 1$ arcmin) and/or faint sources.

In determining the flux correction factor we are able to classify
objects as extended or point-like using the ratio of the King profile
flux correction factor to the point-like flux correction factor
$f_{King}/f_{PS}$ ($f_{King}=s_{true}/s_{detected}$), where
$s_{detected}$ is estimated assuming a King profile, $f_{PS}$ is the
same factor but with $s_{detected}$ estimated assuming a point
source).  If this ratio exceeds some value (not necessarily unity,
because of noise) then we deduce that the source is best fit by an
extended surface brightness profile (through the integral quantities).
This extent criterion has been empirically determined from both these
simulation results and the survey data itself. If the ratio of the flux
correction factors ($f_{King}/f_{PS}$) is {\em greater} than 1.2 then
 $>90$\% of on-axis extended sources with intrinsic extent $r_{c}
 \gtsim 7 $ arcsec will be correctly classified as extended, and
$\ltsim 10$\% of on-axis point-like sources will be mistakenly
classified as extended (from simulations of point-like sources). All
high signal-to-noise off-axis sources with intrinsic extent $r_{c}
\gtsim 20$ arcsec will be correctly classed as extended. In Figure
7(a,b) we summarize the results of applying this classification to the
simulation data.  The fraction of sources classified as extended is
plotted against intrinsic source extent. It is apparent that off-axis
extended sources are more likely to be mis-classified. Given that the
PSF FWHM (at 1 keV) on-axis is 25 arcsec and increases to 58 arcsec
(\markcite{Hasin93b} Hasinger et al 1993b) at 15 arcmin off-axis we
might expect a  degradation of the method, especially at low
signal-to-noise.

The optical follow-up observations we perform for all candidates will
allow us to reliably identify the few expected mis-classified objects.
 To our flux limit of $3.5\times 10^{-14}$ erg s$^{-1}$ cm$^{-2}$ we do
 not expect any extended objects (clusters, groups or galaxies) to have
 core radii of less than $\simeq 20$ arcsec, based on the canonical
 range of physical sizes and luminosities (see Figure 8). Our sample of
 X-ray  extended objects will therefore be complete to this flux limit,
 assuming no evolution in these canonical sources. If, for example,
 clusters of the same luminosity were physically smaller at higher
 redshifts, they might not be classified as extended. However, they
would be identified correctly as cluster candidates from our optical
imaging of point-like X-ray sources.

A visual inspection of a subset of the real data confirms these
results. Of
 the objects classified by eye as extended, 92\% (23 of 25) were
 classified as extended by VTP. Of the objects classified by eye as
point-like, 4\% were misclassified by  VTP. Many of the
misclassifications involved two or more close point-like sources,
separated by a distance similar to the PSF full width half maximum, too
close to be deblended.

\section{Sky coverage} In order to correctly evaluate any statistical
measurements of the survey sample  (such as LogN-LogS, luminosity
functions, extent distribution etc.), the effective sky coverage must
be known. Since the WARPS uses pointed data each field has a different
detection sensitivity to  objects of a given extent and flux, according
to exposure and background. To estimate this, we have combined an
analytic measure of VTP's detection sensitivity with the results of
simulated PSPC data to ensure its validity. On the basis of simulations
we have determined that we can parameterize the criterion for
 for VTP to successfully detect a real source using a definition of the
detected signal-to-noise.  The criterion for source detection is then
approximated as; \begin{equation}
\frac{n_{VTP}-n_{back}}{\sqrt{n_{VTP}}} > 3 \;\;\; ,  \end{equation}
where $n_{VTP}$ is the total number of photons (source and background)
that lie within a radius $r_{VTP}$ which is the equivalent source
radius $\sqrt{A_{VTP}/ \pi}$ where $A_{VTP}$ is the source area within
which the surface brightness exceeds the VTP surface brightness
threshold $\sigma_{VTP}$, which is
 defined by the background ($\sigma_{back}$) and threshold ($f_{min}$); $\sigma_{VTP}=f_{min}
\sigma_{back}$ ($f_{min} \geq 1.0$).  If we assume a King profile as in
Equation 1 then we can write:  \begin{equation} r_{VTP} = \left[
r_{c}^{2} \left[ \left( \frac{\sigma_{VTP}}{\tilde{\sigma_{0}}} \right) ^{-2/3}
-1 \right] \right]^{1/2}\;\;\; . \end{equation}  Here, however, we have
defined the central surface brightness so as to include the background,
$\tilde{\sigma_{0}}=(s_{true}/2 \pi r_c^{2}) + \sigma_{back}$, in units of
counts per unit area. The number of photons detected by VTP should then
be; \begin{equation} n_{VTP}(r_{VTP}) = 2 \pi r_{c}^{2} \tilde{\sigma_{0}}
\left[ 1- \left( 1+ \left(\frac {r_{VTP}}{r_c}\right) ^{2} \right)
^{-1/2} \right ] \;\;\; .  \end{equation}

Instead of explicitly including the full (numerical) PSF we have
included it as a simple `blurring' of the  extent (i.e. the effective extent is
$r_{c}^{eff}=\sqrt{r_{c}^{2} + PSF(\theta)_{\sigma}^{2}}$ where $\theta$ is off-axis angle).  This simplification
appears justified when results are compared with simulations. Using the
exposure and background information (as determined by VTP)  for each of
the 91 fields used in the survey we then integrate the sky coverage
over the radius of each PSPC field, including the PSF variation with
off-axis angle, using the criteria described above to determine the
detection sensitivity (for a given $\sigma_{0}$ and $r_{c}$). The final
result shown in Figure 8 is the combined sky coverage of all fields
used in the survey to the limiting threshold ($f_{min}=1.0$).

In Figure 8 the fractional sky coverage is shown as a function of
projected (intrinsic) extent and intrinsic flux, assuming a King
profile as in Equation 1. The dashed contour denotes  1\% coverage and
indicates that a few deep fields in our survey do
 indeed allow detections to much lower fluxes. Solid contours run in
 steps of 10\% from 10\% to 100\% (with increasing weight). For the
 WARPS-I sample 100\% sky coverage corresponds to 17.2 deg$^{2}$.

  The near-horizontal dashed, dotted and solid curves at the lower
  right in Figure 8 are the loci with varying redshift of
  (respectively) elliptical galaxies (with $L_{x} $(0.5-2 keV) $= 1
\times 10^{42}$ erg/s and effective core radius $r_c = 50 $ kpc),
groups (with $L_{x}$ (0.5-2
 keV)$ = 1 \times 10^{43} $ erg/s, $r_c = 100$ kpc) and clusters (with
$L_{x}$ (0.5 - 2 keV)$ = 5 \times 10 ^{44}$ erg/s, $r_c = 250$ kpc)
($H_{0}=50$, $q_{0}=0$). Redshift is therefore increasing right to left
along these curves.
 The vertical dashed line at $3 \times 10^{-14}$ erg s$^{-1}$ cm$^{-2}$
 (0.5-2 keV) represents the approximate lower flux limit of the survey.
 The redshifts of the three object types at this flux limit are listed
 in Figure 8.  As discussed elsewhere we have chosen a flux limit in
 {\em observed} and uncorrected flux of $\sim 3.5 \times 10^{-14}$ erg
 s$^{-1}$ cm$^{-2}$ (corresponding to a $3 \times 10^{-3}$ ct/s count
 rate). The true intrinsic flux limit is therefore slightly higher.

Features to note from this plot are a) for these canonical classes of
object we always have a sky coverage of $\sim 60$\% or better and b) we
have good sky coverage ($\geq 50$\%) for moderately bright but very
extended sources ($r_{c} > 1$ arcmin) should such low surface brightness
objects exist.

In order to check this calculation we have compared it with the results
of several hundred simulated objects processed by VTP. These were
constructed to have a range of fluxes, extents and off-axis angles. The
full PSF (\markcite{Hasin93b}Hasinger et al 1993b) was used to generate
the final, simulated PSPC fields.  The agreement with simulations is
extremely good and confirms the choice of 3 as a signal-to-noise
detection criterion. Note that at the survey flux limit the lowest
signal/noise of any real source detection in the 91 fields is 4.4, and
most sources have a signal/noise $>6$.

In Figure 9 the WARPS fractional sky coverage is shown as a function of
redshift for three classes of object.  We have nearly complete sky
coverage for clusters with  $L_{x}$ (0.5 - 2 keV)$ = 5 \times 10 ^{44}$
erg s$^{-1}$ and core
 radius $r_c = 250$ kpc out to a redshift of $z \simeq 1$, with 50\%
coverage  at a redshift of $z \sim 1.4$. We have good coverage for
groups (or faint, small clusters) of $L_{x}$ (0.5-2 keV) $=1\times
10^{43}$ erg s$^{-1}$ out to
 $z \sim 0.2$ (80\% coverage) and for galaxies to about $z \sim 0.1$
 (80\%).  We will use this information in evaluating (for example) the
LogN-LogS measured by the WARPS cluster survey. It supports the
discussion in Section 1 on why X-ray selected cluster samples have many
advantages over  optically selected ones. At a redshift of $\sim 1$ the
number density of galaxies on the sky means that the contrast of any
cluster is greatly reduced in the optical, whereas the X-ray contrast
is still high.

\section{A comparison with conventional detection methods}

An important result of this work is the demonstration that a method
such as VTP can detect sources (especially extended, low surface
brightness sources) sometimes missed by conventional source detection
algorithms (which are known to be less suited to detection of extended
objects). Previous work (\markcite{Ebeli96b} Ebeling et al 1996b) has
shown that VTP detects more sources than the Standard Analysis Software
System (SASS, \markcite{Voges92} Voges et al. 1992) in the Rosat All
Sky Survey (RASS).  In order to better quantify this difference in
source detection efficiency for the PSPC fields we have compared our
VTP method with a standard sliding cell method. This sliding cell
algorithm is publicly available as part of the XIMAGE data analysis
package\footnote{\tt
http://heasarc.gsfc.nasa.gov/docs/xanadu/ximage/node1.html} and is
referred to as the DETECT algorithm. Briefly, DETECT divides the image
field into boxes to make local estimates of the background (rejecting
those which deviate significantly from Poisson expectations) and then
forms a global background estimate (rejecting the tails of the
Gaussian  distribution). The sliding cell size is chosen to optimize
signal-to-noise, and corrections are made for dead time, vignetting
effects, and the fraction of the source counts that fall outside the
box where the net counts are estimated. Count rate errors are
included.

We have run DETECT on a subset of our survey fields, in the energy band
0.5-2 keV, and using the default parameters (such as threshold, source probability etc.). In
Figures 10,11,12 and 13,14,15 some examples of the problems encountered
by conventional techniques are shown.

In Figure 10, the raw photon map for this field is shown, with the
Voronoi cells plotted. Many potential sources are apparent to the eye
in this crowded field.  In Figure 11 the results of running VTP at the
lowest threshold (1.0) are presented. Those sources with background
corrected counts of less than 20 photons are labeled with `X'. The
vertical arrows indicate sources found by DETECT. It is clear that
there are several places where DETECT and VTP differ. VTP sources 1, 3
and the photons to the North of source 15 are not
 seen as enhancements by DETECT. In Figure 12 the picture is clearer at
 the higher VTP threshold of 1.3. Source 19 which was previously
 blended into source 15 in Figure 11 is now distinct (also see Figure
 10) and is a good candidate (given the lower threshold observation)
 for an extended source, but is completely missed by DETECT.

In Figures 13,14,15 the results for another field are shown in the same
format as for Figures 10,11,12. In Figure 14, VTP sources 1, 21,26 and
27 have no DETECT counterparts. In Figure 15, at the higher VTP
threshold of 1.3 we can see that these sources are still significant
(now numbered 1, 21, 27 and 28 respectively and are good candidates for
extended sources (see also Figure 13).

As a further test, DETECT was run with a greatly reduced threshold for
source detection in order to see if it could indeed detect the sources
seen by VTP at lower surface brightness. Even when the threshold was
lowered to a point at which more than twice the original number of
detections were obtained, many VTP sources remained undetected.
Furthermore, at this low threshold many of the DETECT sources were
clearly spurious, indicating that even if the DETECT parameters could
be altered to find the VTP sources it would be difficult to distinguish
these from the false detections.

While those fields in Figures 10--12 and 13--15 were chosen
specifically to demonstrate the ability of VTP to detect extended
sources missed by conventional methods they are not atypical. To
quantify the differences we
 have used a subset of ten fields from the survey across a range of
exposure times and backgrounds. Using the VTP and DETECT algorithms we
have made two flux limited samples using this subset (to the same flux
limit as  WARPS-I).  To ensure the robustness of the VTP detections we
used the results of the 2nd surface brightness threshold only
($f_{min}=1.3$).  As demonstrated in fields 700114 and 600520 above the
DETECT algorithm finds
 fewer objects to a given flux limit. In terms of the number counts of
all objects with flux
 $\gtsim 3.5\times 10^{-14}$ erg s$^{-1}$ cm$^{-2}$ (0.5-2 keV) VTP
 finds a number per unit area of $\simeq 19.7\pm 2.8$ deg$^{-2}$
 compared to DETECT which finds $\simeq 18.1\pm 2.7$ deg$^{-2}$ from
 the same data. This difference in counts is almost exactly that which
 would be expected at this flux limit if the additional sources
 detected by VTP are in fact extended (based on the results of
 \markcite{Rosat95} Rosati et al 1995 who use a wavelet method).

While such sliding cell detection methods can be altered to improve
their efficiency for extended objects it is clear that in the worst
cases they can miss almost {\em all} fainter extended objects in a flux
limited survey. This has profound implications for any survey of faint
extended objects which does not use a detection method sensitive to low
surface brightness objects. For example, the recent results of
\markcite{Casta95} Castander et al (1995) (who also survey ROSAT PSPC
fields) indicate that to the same flux limit as used here they detect
approximately one object per deg$^{2}$ {\em less} than this present
work or the work of \markcite{Rosat95} Rosati et al (1995). This suggests that
there may indeed be a difference in the sensitivity of the X-ray
detection methods used.

\section{Results and Discussion}

The WARPS cluster survey has a current sky coverage of $17.2$
 deg$^{2}$. Using the VTP  method this sample (WARPS-I) contains 298
 objects (uncorrected for sky coverage) with detected flux $> 3.5\times
 10^{-14}$ erg s$^{-1}$ cm$^{-2}$ (0.5-2 keV). There are 58 extended
 sources according to the criterion used in Section 3.3. The number of
 misclassified point-like sources is estimated to be $\approx 5$\% of
 the total number of point-like sources (i.e. about 12 sources), giving
 a possible range of 2.8--4.0 extended objects deg$^{-2}$. The maximum
 number of misclassified point-like sources, assuming they all have
detections of low signal-to-noise (which, in reality, they do not) is
10\%, which gives a lower limit of 2.0 extended objects deg$^{-2}$.
  The combination of a detection method which is unbiased by shape and
surface brightness with a rigorous optical follow-up results in an
extremely well selected and quantified catalogue well suited to
measuring properties such as cluster evolution.

A simple test of the effectiveness of our approach in identifying
extended objects can be made by studying the results obtained for known
high redshift clusters. In Table 1 below, 4 known clusters spanning
redshifts from z$=0.13 $ to 0.66 were detected by VTP in the PSPC
fields and flagged as being extended.
The cluster J1836.10RC was detected twice in separate pointings of
exposure 21 ksec and 23 ksec. The raw VTP flux was 2.8x10$^{-14}$ erg
cm$^{-2}$ s$^{-1}$ (0.5-2 keV) (55 photons), and the flux variation
between observations was 20\% or 1.2$\sigma$, showing a repeatability
of the VTP measurements  within that expected from Poissonian
statistics.

We find a sky density of 2.1--3.1 $(\pm 0.4)$ extended objects
deg$^{-2}$ to an intrinsic (as opposed to detected) flux limit of $ 4.5
\times 10^{-14}$ erg s$^{-1}$ cm$^{-2}$. This is in good agreement with
the survey of \markcite{Rosat95} Rosati et al. (1995), although their
initial sample contained only $10$ extended objects and our numbers are
based on a lower limit of $ 40$  extended objects. The survey of
\markcite{Casta95} Castander et al (1995) which had a similar survey
area and a flux limit of $3 \times 10^{-14}$ erg s$^{-1}$ cm$^{-2}$
found a sky density of $0.9\pm 0.25$ extended object deg$^{-2}$. There
is clearly a discrepancy between these observations.  We have
demonstrated (in agreement with previous works,
\markcite{Ebeli96b}Ebeling et al 1996b) that the VTP method detects
more objects than conventional methods, which were designed with point
source detection more in mind.  In the extreme case, based on an exact
comparison to our flux limit ($\sim 3.5\times 10^{-14}$ erg s$^{-1}$
cm$^{-2}$) VTP detects some 2 objects deg$^{-2}$ {\em more} than a
sliding cell method. This indicates that a large fraction of extended
objects may be missed at this flux limit.  Until we have redshifts for
these low surface brightness objects we cannot estimate the effect of
their exclusion from previous surveys; however we suggest that counts
of low luminosity, moderate redshift clusters obtained with non-optimal
detection algorithms should not be considered reliable.

\subsection{Surface brightness limits and LogN-LogS}

While often quoted in optical surveys the limiting surface brightness
is not generally presented in X-ray surveys. In Figure 16 we present
the distribution of surface brightness limits for the WARPS cluster
survey fields used in this present work. This is simply determined from
the observed background counts in the fields and the VTP surface
brightness thresholds.  The vertical dot-dashed and dashed lines
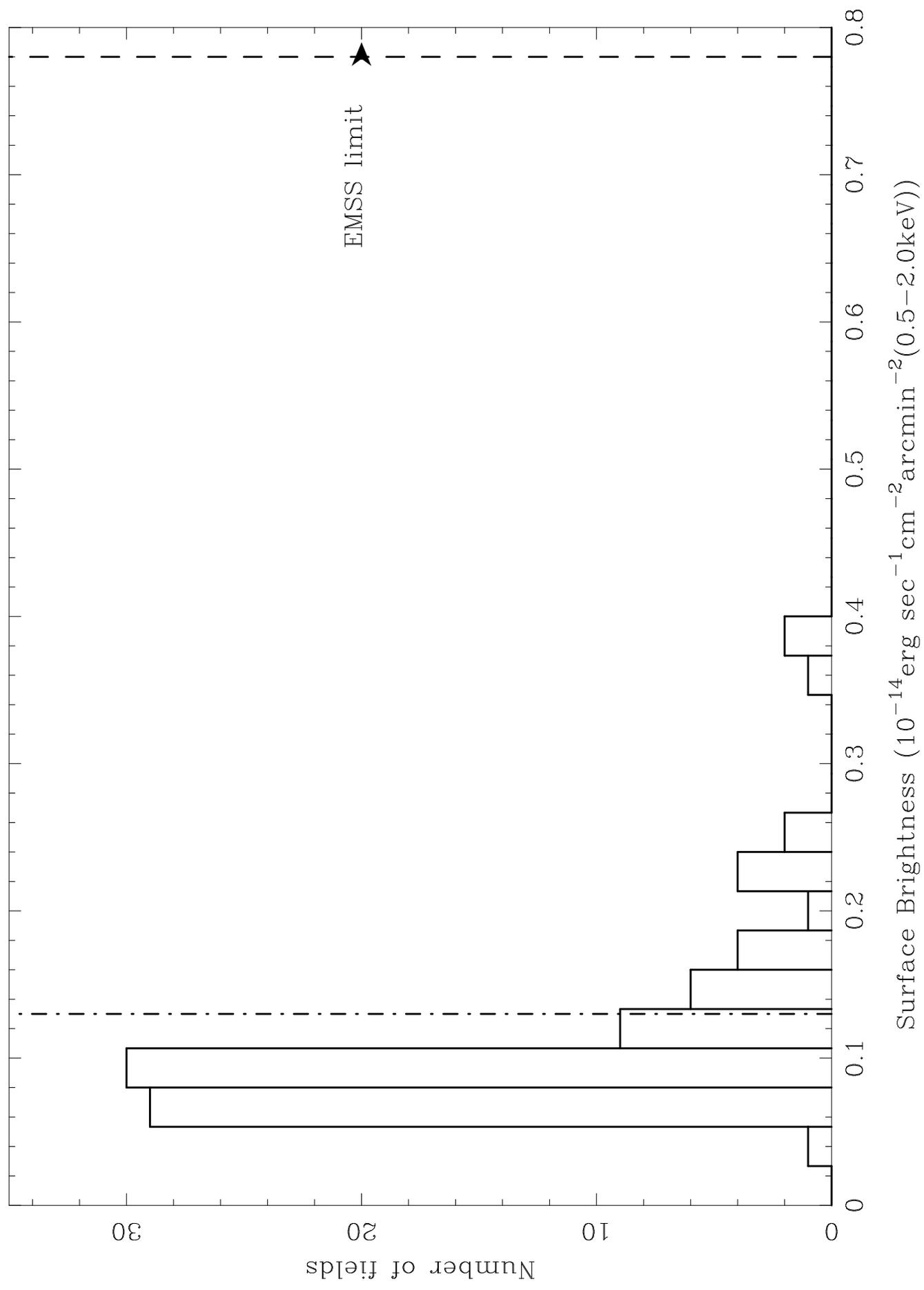
represent a comparison of the WARPS surface brightness limits with that
of the EMSS respectively.  For example, a limiting surface brightness
of 1.3$\times$10$^{-15}$ erg s$^{-1}$ cm$^{-2}$ arcmin$^{-2}$ (0.5-2
keV) (dot-dashed line in Figure 16) in a typical ROSAT PSPC field with
a background  level of 3 ct/arcmin$^{2}$ in 15 ksec exposure
corresponds to a two sigma excess over the background in a
2.4$\times$2.4 arcmin cell (the EMSS detection cell).  To compare this
with the limiting surface brightness in the EMSS, parameters for a
typical EMSS field were taken from \markcite{Gioia90}Gioia et al
(1990).  For an Einstein IPC exposure of 3.5 ksec and the same
2.4$\times$2.4 arcmin cell size containing 9 background counts, a two
sigma excess was given by a surface brightness of 1.4$\times$10$^{-14}$
erg s$^{-1}$ cm$^{-2}$ arcmin${-2}$ (0.3-3.5
 keV) or 7.8$\times$$^{-15}$ erg s$^{-1}$ cm$^{-2}$ arcmin${-2}$ (0.5-2
 keV) (dashed line in Figure 16), using an  EMSS (0.3-3.5 keV) to ROSAT
 (0.5-2 keV) band conversion factor of 0.55, appropriate for a power
 law spectrum of energy index 1 or a thermal spectrum of temperature
 kT=6 keV. The typical WARPS surface brightness limit is therefore
 $\approx 6$ times lower than that of the EMSS. We then expect to
 detect more flux from  low surface brightness emission than in (for
 example) the EMSS.

In Figure 17 we present the first number count results for the WARPS-I
cluster survey. The LogN-LogS relationship is shown for the 91 PSPC
fields and 298 objects analysed at the time of writing (a total sky
coverage of $ 17.2$ deg$^{2}$). The 0.5-2 keV count rates were
converted to fluxes using a constant factor of $1.15 \times 10^{-11}$
erg s$^{-1}$ cm$^{-2}$ (ct s$^{-1}$)$^{-1}$. This conversion factor is
accurate to within 7\% for a thermal \markcite{Raymo77}Raymond-Smith
(1977) spectrum of temperature kT$=$1-20 keV, abundances 0.25-1 times
cosmic abundance, and column densities $1\times 10^{20}$ cm$^{-2}$ -
$1\times 10^{21}$ cm$^{-2}$. These counts include {\em all} sources
(extended or pointlike). The counts have been corrected for sky
coverage assuming all sources to be pointlike, this will result in an
 underestimate of the number counts. The raw flux data points therefore
represent a {\em lower limit}. The initial results for the corrected
flux points are shown (with pointlike sky coverage correction) to
demonstrate the size of the flux correction.  We have also plotted the
LogN-LogS relationships found by \markcite{Hasin93a}Hasinger et al
(1993a) and \markcite{Brand94}Branduardi-Raymont et al (1994)
(hereafter BR94), from ROSAT data and Gioia et al (private communication
in BR94) from Einstein EMSS data converted to the ROSAT band. The
curves represent the LogN-LogS relationships of all sources, as
detected with PSF-based detection algorithms.

We see a clear excess at the brighter end compared with other
measurements. There are two possible explanations; first, that the VTP
method simply detects more objects because it is sensitive across a
greater range of surface brightness than other detection methods.
Second, that the observed VTP flux is higher for bright extended
objects because it includes the signal from the lower surface
brightness parts of sources. A combination of these two effects is also
likely.
 This result suggests that a single power law is insufficient to
 describe the counts over this range in flux ($\sim 3-10 \times
 10^{-14}$ erg s$^{-1}$cm$^{-2}$ (0.5-2  keV)).

If we assume that the deviation from other measurements is due to the
improved detection of extended sources (since all detection methods
should be essentially equal in the detection efficiency of point-like
sources) then we are detecting a population of low surface brightness
sources with a density of $\sim 1 $deg$^{-2}$ at a flux greater than
$\sim 1 \times 10^{-13}$ (with some confidence since we actively avoid
fields with known, bright, clusters as targets).  Alternatively, if we
are not detecting an additional population but rather a greater flux
for previously detected objects (presumably extended ones, using the
above reasoning) then this amounts to a $\gtsim 20$\% increase in
measured flux for objects with flux greater than $10^{-13}$.  We have
made a preliminary investigation of those sources contributing to these
higher flux counts which suggests that the excess in our LogN-LogS is
indeed due in part to an improved detection of the low surface
brightness component of brighter, extended sources.

\section{Summary and Conclusions}

The ability to test cosmological models through detailed studies of
galaxy clustering calls for well selected surveys which span the entire
range of cluster types. Since the dominant luminous mass component in
such systems is hot, X-ray emitting gas this offers an ideal way of
selecting objects, it is also free of many of the uncertainties in
optically derived catalogues. In order to extend the currently probed
section of the cluster population to lower luminosities and surface
brightnesses and higher redshifts we must use general, unbiased source
detection methods such as VTP.

The lower limit LogN-LogS of all sources in WARPS-I shows a significant
excess in counts for fluxes greater than $\sim 10^{-13}$erg s
$^{-1}$cm$^{-2}$ (0.5-2 keV) compared to previous works. This excess is
consistent with the combined point-like and extended source counts of
\markcite{Rosat95}Rosati et al (1995) and appears to be  due to an
improved detection of the flux associated with extended sources. We
have determined our surface brightness limit to be $\sim 1\times
10^{-15}$ erg s$^{-1}$cm$^{-2}$ arcmin$^{ -2}$ (0.5-2 keV) which is
approximately 6 times lower than that of the EMSS, in agreement with
our excess counts. On the basis of a pure X-ray classification we
determine a sky density of 2.1--3.1 $(\pm 0.4)$ extended objects
deg$^{-2}$ in the present sample with intrinsic flux $>4.5 \times
10^{-14}$erg s$^{-1}$ cm$^{-2}$ (0.5-2 keV).

Our survey has complete sky coverage (17.2 deg$^{2}$) for rich clusters
($L_{X} = 5\times 10^{44}$erg s$^{-1}$) to a redshift of 1. At our flux
limit of $ 3.5\times 10^{-14}$erg s$^{-1}$cm$^{-2}$ we can detect (with
$\sim 50-60$\% sky coverage) rich clusters to z=1.16, groups or poor
clusters to z=0.23 and
 individual galaxies to z=0.08, assuming the canonical sizes and
 luminosities of such objects. We have confirmed our detection of
 extended objects at high redshift by comparison with known clusters
 (Table 1). Furthermore we find that using the appropriate exposure
maps for the ROSAT PSPC fields can improve flux estimates by as much as
15\% compared to a uniform correction and can reduce the variations
between flux estimates in different fields (if only a standard
correction were applied) by as much as 15\%.

We have demonstrated that VTP detects more low surface brightness
emission than conventional PSF/sliding cell based methods. In the worst
cases such methods can miss a large fraction of fainter extended
objects in a flux limited survey. This has profound implications for
any survey of faint extended objects which does not use a detection
method sensitive to low surface brightness objects. For example, the
recent results of \markcite{Casta95}Castander et al (1995) (who also
survey ROSAT PSPC fields) when compared
 with the results in this paper or of \markcite{Rosat95}Rosati et al
 (1995) suggest a significant difference in the detection sensitivity
 of the methods used.

Work in progress will address the issues of the cluster LogN-LogS,
evolution in the cluster XLF, cluster morphologies and optical
classifications using the WARPS-I survey.

{\bf Acknowledgements}

The data used in this work have been obtained from the High Energy
Astrophysics Science Archive Research Center (HEASARC) at NASA Goddard
Space Flight Center (http address).  CAS and LRJ acknowledge Regular
and Senior NRC Research Associateships respectively. We acknowledge
discussions with Nick White and Richard Mushotzky and thank Nick White,
Lorella Angelini and Paolo Giommi for producing WGACAT, on which early
work was based.

\newpage

\begin{table}[h] \caption{Known clusters detected as extended X-ray
sources by VTP} \label{tbl-1} \begin{center} \begin{tabular}{llll}
Cluster & Reference & Redshift & Note\\ \tableline
 & & & \\ Pavo & Griffiths et al 1992 & 0.13 & In Einstein deep survey
field \\ J1836.10RC & Couch et al 1991 & 0.275 & \\ 0055-279 & Roche et
al 1995 & 0.56 & In SGP field \\ F1767.10TC & Couch et al 1991 & 0.664
& \\ \end{tabular} \end{center} \end{table}

\newpage

\bf
\noindent{Figure captions:}
\rm

\noindent{\bf Figure 1:} The 91 ROSAT pointings selected for this
initial survey (Aitoff projection in Galactic coordinates). Points are
weighted by exposure time. The dotted horizontal lines delimit
$|b|=20^{\circ}$, the hatched line is the equatorial coordinate system
equator ($\delta=0^{\circ}$).

\noindent{\bf Figure 2:} The Voronoi tessellation of a typical ROSAT
PSPC photon distribution (the target is a star, HD 173524). Photons are
shown as points, the tessellation cells occupy the inner 18 arcmin
radius of the field. Sources are immediately apparent to the eye as
clusters of small cells.

\noindent{\bf Figure 3:} The source photons and sources identified by
VTP at threshold 1.0 in the field shown in Figure 2. Heavy points
indicate VTP source photons, sources are labelled numerically. The
vertical arrows mark the positions of sources from WGACAT (which uses a
conventional detection
 algorithm, in a broader energy band). Note that some
 VTP sources (such as 2 and 6) are clearly potential blends.

\noindent{\bf Figure 4:} The source photons and sources identified by VTP at threshold
1.7 in the field shown in Figure 2.  Note that sources 2 and 6 at
threshold 1.0 have now been resolved out into sources 2,6,8,9,10  and
11,13,15 respectively.

\noindent{\bf Figure 5 a,b:} The mean ratios of detected and corrected
count rates to the true count rate of low signal-to-noise ($s/n < 8$)
simulated sources plotted against (a) source extent (arcsec) and (b)
source count rate. Each point is the mean ratio of the Monte Carlo
realisations within a simulation of a given set of parameters.
 Open symbols represent raw, detected count rates, filled symbols
represent count rates corrected as described in Section 4. Circular
points denote sources simulated to be on-axis, triangular points denote
sources simulated to be 15 arcmin off-axis. Error bars show the
 $1\sigma$ scatter.

\noindent{\bf Figure 6:} As in Figure 5(a,b) but for high
signal-to-noise sources ($s/n > 8$).

\noindent{\bf Figure 7 a,b:} The fraction of simulated sources
recognized as extended (with an extended source criterion that
$f_{King}/f_{PS} \geq 1.2$) is plotted against intrinsic source extent.
In a) the results for low signal-to-noise sources ($s/n<8$) are plotted
in b) the results for high signal-to-noise sources ($s/n>8$) are
plotted. Circular symbols represent the results for on-axis sources,
triangular symbols represent sources at 15 arcmin off-axis. The symbols
are scaled according to the number of source photons.

\noindent{\bf Figure 8:} The sky coverage offered by the 91 fields used
in this initial survey. Sky coverage is plotted as a function of
intrinsic flux and intrinsic, projected extent (assuming a King
profile).  Contours are drawn at percentage of total survey area. The
dashed contour is at a level of 1\%, subsequent solid contours (of
increasing weight) are 10\%,20\%... to 100\%. The near horizontal {\it
dashed, dotted} and {\it solid} curves at the lower right are the loci
with redshift of  (respectively) {\em elliptical} galaxies ( $L_{x}$
(0.5-2 keV)$ = 1 \times 10^{42}$ erg/s, $r_c = 50 $ kpc), {\em groups}
( $L_{x} $(0.5-2 keV)$ = 1 \times 10^{43}$ erg/s, $r_c = 100 $ kpc) and
{\em clusters} ($L_{x} $(0.5 - 2 keV)$ = 5 \times 10 ^{44} $erg/s, $r_c
= 250$
 kpc) ($H_{0}=50$, $q_{0}=0$). Redshift is therefore increasing right
 to left and is different for each curve.
 The vertical dashed line at $3.5 \times 10^{-14}$ erg sec$^{-1}$
 cm$^{-2}$ (0.5-2 keV) represents the approximate lower flux limit of
 the survey. The redshifts of the three object types at this flux limit
 are listed.

\noindent{\bf Figure 9:} The WARPS sky coverage as a function of redshift
for three classes of objects: Elliptical galaxies (with $L_{x}$ (0.5-2
keV)$ = 1 \times 10^{42}$ erg s$^{-1}$ and effective core radius $r_c =
50$ kpc), Groups ($L_{x}$ (0.5-2 keV)$ = 1 \times 10^{43}$ erg
s$^{-1}$, $r_c = 100$ kpc) and Clusters (with $L_{x} $(0.5 - 2 keV)$ =
5 \times 10 ^{44}$ erg s$^{-1}$,  $r_c = 250$ kpc). Light and heavy
curves represent cases in which $q_{0} =0$ and 0.5 respectively. The
Hubble constant $H_{0}$ (km/s/Mpc ) is 50 in all cases.

\noindent{\bf Figure 10:} The raw PSPC photons and Voronoi cells for field 700114.

\noindent{\bf Figure 11:} The same field (700114) as in Figure 10. VTP
source detections are plotted as heavy points, detections are made to
the lowest surface brightness threshold used in the survey. Those VTP
sources with background corrected photon counts less than 20
 are labelled with `X'. Vertical arrows indicate the source detections
 of the sliding window algorithm.

\noindent{\bf Figure 12:} As in Figure 11, but the surface brightness
threshold used by VTP is now 1.3.

\noindent{\bf Figure 13:} The raw PSPC photons and Voronoi cells for field 600520.

\noindent{\bf Figure 14:} The same field (600520) as in Figure 13. VTP
source detections are plotted as heavy points, detections are made to
the lowest surface brightness threshold used in the survey. Those VTP
sources with background corrected photon counts less than 20
 are labelled with `X'. Vertical arrows indicate the source detections
 of the sliding window algorithm.

\noindent{\bf Figure 15:} As in Figure 14, but the surface brightness
threshold used by VTP is now 1.3.

\noindent{\bf Figure 16:} The distribution of surface brightness limits
in the WARPS cluster survey fields. The spread in values reflects the
spread in background counts in the PSPC fields. The vertical dot-dashed
line at a surface brightness of 1.3$\times 10^{-15}$ corresponds to a
two sigma surface brightness detection in a square cell (see text) for
a typical ROSAT field. The vertical dashed line to the right represents
the equivalent significance typical surface brightness detection limit
of the EMSS, adjusted to the 0.5-2 keV band. It is $\sim 6$ times
higher than the mean WARPS limit.

\noindent{\bf Figure 17:} The LogN-LogS of all sources in the initial
WARPS survey  (0.5-2 keV). The number counts for raw (corrected only
for backgrounds) fluxes (plotted as filled symbols) have been
corrected for sky coverage assuming zero object extents, these data
points should therefore be considered as lower limits for the survey.
The number counts for corrected fluxes (open symbols) have also been
corrected for sky coverage assuming zero extent objects.  The curves
are the results of earlier works, Hasinger et al (ROSAT PSPC data)
1993a (solid curve), BR (ROSAT data) 1994  (dashed curve) and the
Einstein Medium Sensitivity Survey (EMSS) (Gioia et al, private comm.
BR94) (dot-dashed curve). Error bars are displayed on only two data
points for illustration, since these are integral quantities the errors
are not independent.  The faintest point for the corrected fluxes is
not plotted since the survey limit in detected flux excluded sources
which would otherwise have been moved into this bin from fainter fluxes
by the flux correction.

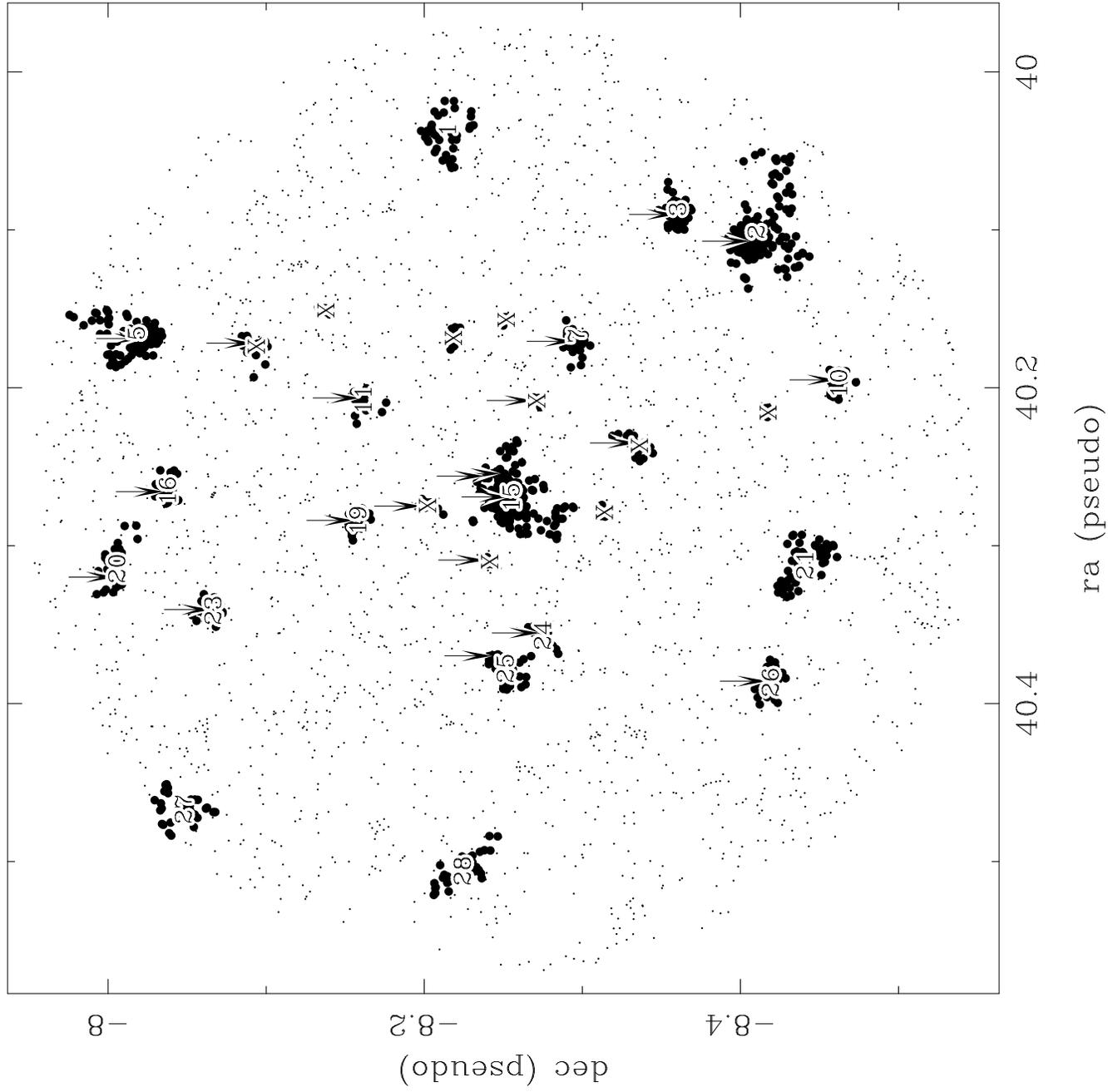
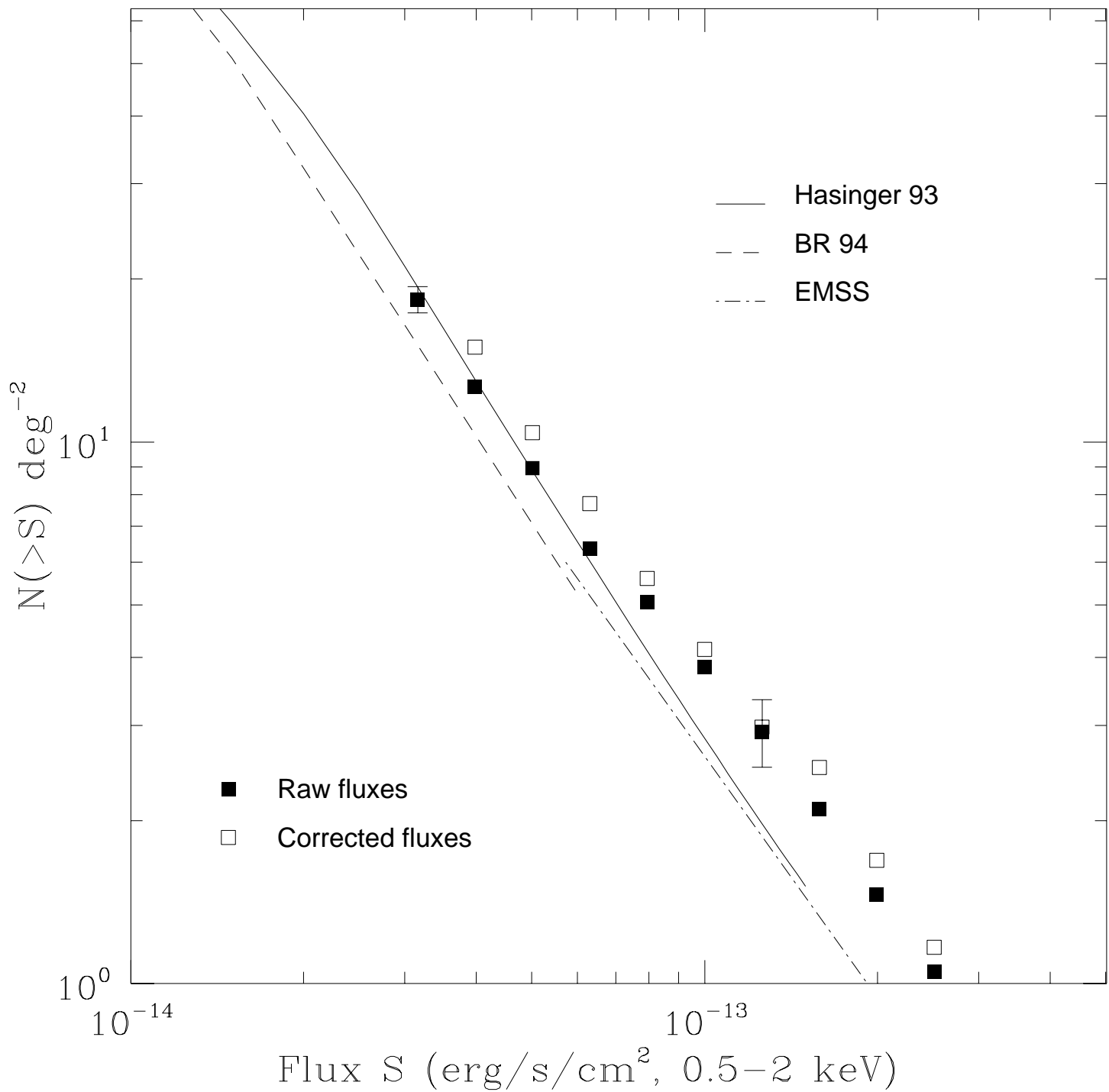
\end{document}